\documentclass[a4paper,12pt]{article}
\usepackage{jinstpub} 
\usepackage{lineno}
\usepackage[table]{xcolor}
\usepackage{subcaption}
\usepackage{colortbl}
\usepackage{caption}
\usepackage{wrapfig}
\definecolor{headerblue}{RGB}{0, 102, 204}
\definecolor{rowgray}{gray}{0.9}

\title{\boldmath CosmicWatch: The Desktop Muon Detector (v3X) }

\author[a]{Spencer N. Axani$,^{1,}$\note{Corresponding author.}}
\author[a]{Masooma Sarfraz,$^{2,}$\note{Lead author. Contributed significantly to the analysis and writing.}}
\author[a]{Miles Garcia,}
\author[a]{Collin Owens,}
\author[]{Katarzyna Frankiewicz,}
\author[b]{Janet M. Conrad,}

\affiliation[a]{Department of Physics and Astronomy, Bartol Research Institute, University of Delaware,\\ 
210 South College Ave, Newark, DE 19716, USA}

\affiliation[b]{Department of Physics, Massachusetts Institute of Technology,\\
77 Massachusetts Ave., Cambridge, MA 02139, USA}

\vspace{0.5cm}

\emailAdd{saxani@udel.edu}

\abstract{
The CosmicWatch Desktop Muon Detector (v3X) is a compact, low-cost, and portable device designed for detecting ionizing radiation, including cosmic-ray muons. Building on previous iterations, the v3X introduces significant hardware and firmware improvements that enhance sensitivity, usability, and data acquisition capabilities. The detector integrates a plastic scintillator and silicon photomultiplier (SiPM), custom designed electronics for signal processing, onboard data storage, OLED display, environmental sensors, and USB connectivity. With a total component cost under \$100 and a build time suitable for high school students, the v3X is ideal for education, outreach, and introductory research applications in particle and astroparticle physics. This paper details the design, performance, and potential use cases of the v3X, supported by example measurements demonstrating its functionality.}


\begin{document}
\maketitle
\flushbottom
\newpage

\begin{figure}[t] 
    \centering
    \includegraphics[width=15cm]{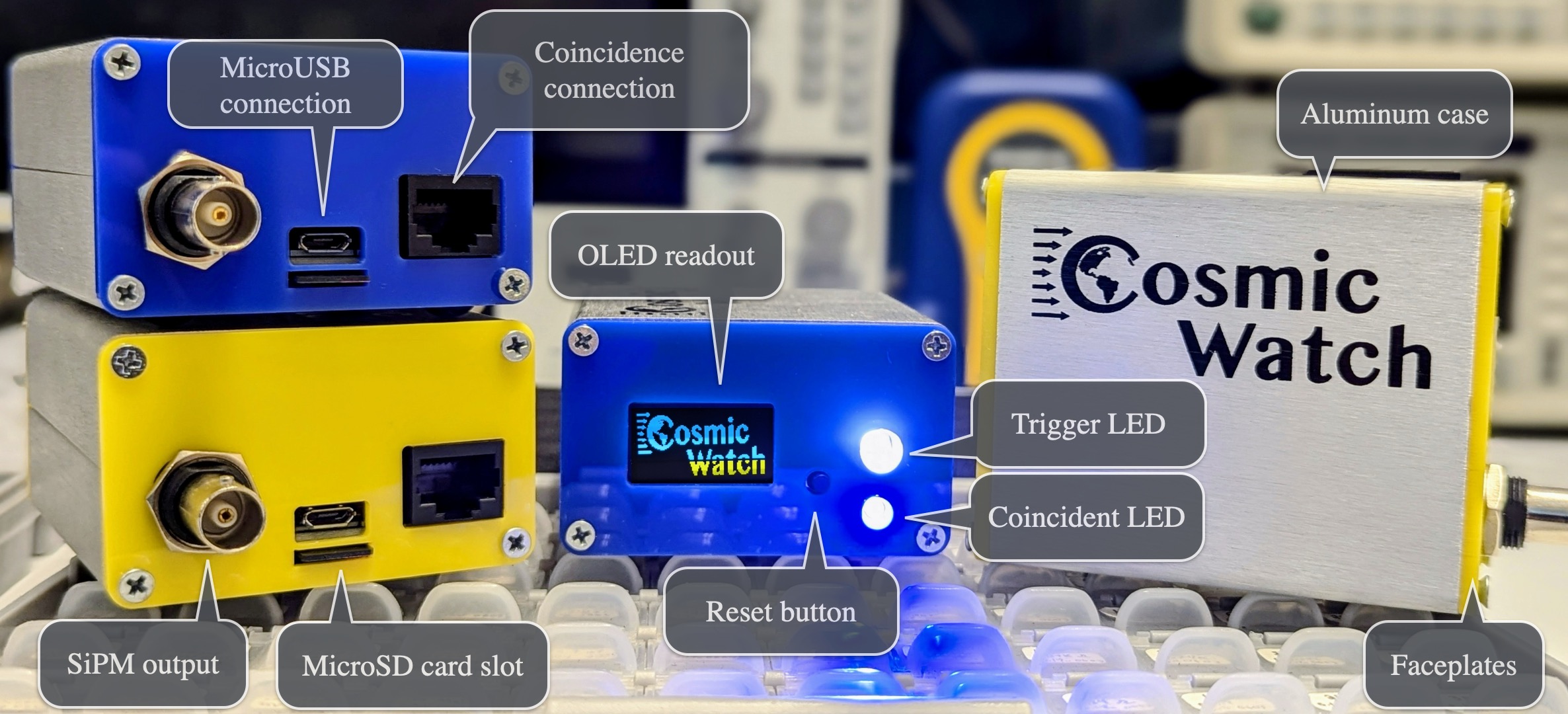}
    \caption{ An array of the CosmicWatch Desktop Muon Detectors v3X. The front and rear faceplates provide access to the SiPM output, microSD card slot, microUSB connection, and coincidence ports. Additional elements include an OLED display, reset button, trigger LED, and coincidence LED.}
    \label{fig:cosmicwatch}
\end{figure}

\section{Introduction}
\label{sec:intro}
\subsection{CosmicWatch}
CosmicWatch\footnote{\url{https://sites.udel.edu/cosmicwatch/}}~\cite{Axani_2018,axani2019,axani2017desktop} is an University of Delaware (UD)-lead outreach program aimed at introducing students to particle, nuclear, and astro-particle physics by providing educational material and support for building particle detectors and making physics measurements. 
\color{black}
The original program, initiated at the Massachusetts Institute of Technology (MIT) and Polish National Center for Nuclear Research (NCBJ) in 2017, has substantially grown and changed, resulting in novel uses for the detectors, including calibration instruments for neutrino detectors~\cite{Calibrationtool}, spacecraft instrumentation, education, and  outreach~\cite{CW_lead_attenuation}\cite{CW_airshower}\cite{airshower_analysis}\cite{collaborationhighschoolsjapan}. It is also substantially used as a versatile platform for teaching high-school, undergraduates, and graduate students practical skills such as soldering, micro-controller unit (MCU) programming, introducing fundamental physics concepts, and giving hands-on experience with particle detection techniques. 

\subsection{The Desktop Muon Detector (v3X)}

This paper presents the v3X Desktop Muon Detector, shown in figure~\ref{fig:cosmicwatch}, with significant improvements over the previous models, v1~\cite{axani2017desktop} and v2~\cite{Axani_2018,axani2019}. These improvements expand the physics capabilities of the detectors, enhance the performance, and add new functionalities, while keeping the total cost of a detector around \$100. A comprehensive list comparing the performance between the two generations of detectors is shown in table~\ref{tab:specifications}. All material pertaining to the v3X detector can be found in the GitHub repository:  

\begin{center}
\small{\url{https://github.com/spenceraxani/CosmicWatch-Desktop-Muon-Detector-v3X}}
\end{center}

\begin{table}[b]
    \centering
    \footnotesize  
    \begin{tabular}{rcl}
        \hline\hline
        \textbf{CosmicWatch v2} & \textbf{Specification} & \textbf{CosmicWatch v3X} \\
        \hline
        ATmega328P  & Processor & \textbf{Cortex-M0+} \\
        Single Core & Number of cores & \textbf{Dual Core} \\
        16 MHz  & Clock frequency & \textbf{133\,MHz}  \\
        32 KB & Flash memory &\textbf{ 2\,MB} \\
        2 KB & RAM & \textbf{264 KB} \\
        \textbf{0.27 W} & Power consumption & 0.5 W \\
        $<$8-bit & ADC resolution & \textbf{12-bit} \\
        15\,Hz & Maximum event rate & \textbf{700\,Hz} \\
        0.5 mV RMS & Noise level & \textbf{0.1\,mV RMS} \\
        Yes & microSD Card & \textbf{Yes} \\
        Yes (w/o SD card) & OLED display & \textbf{Yes} \\
        Yes & Temperature sensor & Yes \\
        No & Pressure sensor & \textbf{Yes} \\
        No & Accelerometer \& gyroscope & \textbf{Yes} \\
        No & Buzzer & \textbf{Yes} \\
        $25\,\text{cm}^2$ & Effective area & $25\,\text{cm}^2$ \\
        50 ms & dead time per event & \textbf{400$\,\mathbf{\mu}$s} \\
        \textbf{106 g} & Dry weight & 110 g \\
        $36\,\text{mm}^2$ & Photocathode area & $36\,\text{mm}^2$ \\
        No & Citizen science support & \textbf{Yes} \\
        12\,mV & Min. trigger threshold & \textbf{4\,mV} \\
        4.5$\times 10^{-4}$\,Hz & Accidental coincidence rate & \textbf{4.5$\times \mathbf{10^{-5}}$\,Hz} \\
        \$100 & Approximate cost & \$100 \\
        \hline\hline
    \end{tabular}
    \caption{Specification comparison between CosmicWatch v2 and v3X detectors. Bold entries indicate the specification with superior performance.}
    \label{tab:specifications}
\end{table}

The updated analog electronics reduce noise, linearize the electronics, and enhance the signal fidelity. The utilization of a dual-core CPU offloads the computationally intensive functions to a second core, reducing the average dead-time by nearly two orders of magnitude. A hardware trigger enables deterministic sampling, which improves the resolution of the detector and analog logic for coincidence detection reduces the accidental coincidence detection by nearly two orders of magnitude. Additional digital peripherals, such as a temperature and pressure sensor (BMP280), and an accelerometer and gyroscope (MPU-6050), provide additional information to help with data analysis. 

With the substantial improvements above, the detectors retain their original functionality, which allows for particle identification (PID), directionality, a measurement of energy deposition, and precise timing; the combination of which enables them to be capable of exploring various physics topics (see Sec.~\ref{sec:measurements}). 

Furthermore, the detector’s modular architecture readily accommodates future upgrades. With minimal modifications it can be reconfigured as a gamma‐ray spectrometer or linked to an expansion module: including GPS receivers, real‐time clocks, magnetometers, WiFi and Bluetooth communication, and environmental sensors. This flexibility lays the groundwork for a planned citizen‐science network, enabling volunteers to map Earth’s ionizing radiation and, ultimately, to reconstruct cosmic‐ray air showers in distributed arrays of detectors. 

Finally, variants of the v3X have already been deployed in diverse contexts from payloads aboard sounding rockets and the International Space Station through NASA's TechRise Student Challenge~\cite{Techrise}, on satellite-based cosmic-ray monitoring using federated learning frameworks. The design has also found applications in classroom education, at MIT Junior Lab, UD Advanced Lab, Cornell undergraduate lab,\footnote{Internal use in undergraduate and graduate instructional labs 8.13 at MIT (2020-2025), UD (2023-2025) PHYS646, and Cornell University (2024-2025)}, and interactive science-art installations, demonstrating its broad utility and adaptability.

\subsubsection{Principle of operation}

When a charged particle (e.g., an electron, proton, or muon) passes through the $5\times5\times1\text{ cm}^3$ plastic scintillator slab, it excites the scintillant molecules, which then de-excite by emitting photons isotopically along the particle’s track \cite{Plastic_scintillator}. A portion of this light is collected by an $6\times6$~mm$^2$ silicon photomultiplier (SiPM), optically coupled to the scintillator with optical gel and/or a silicone pad. The SiPM converts each detected photon into an electrical pulse, the amplitude (or pulse height) of which is proportional to the light intensity.

\begin{figure}[t]
    \centering
    \includegraphics[width=\textwidth]{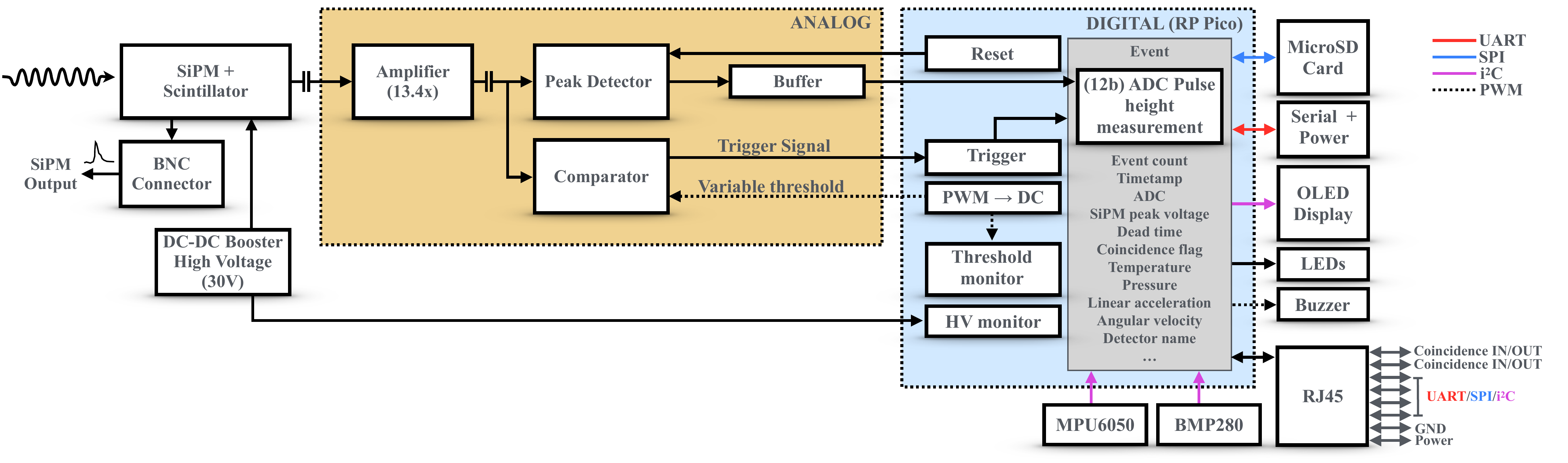}
    \caption{Block diagram of the main board. The analog section shows the amplification, peak detection, as well as the comparator. The digital section illustrates the communication between peripherals as well as monitors and controls.}
    \label{fig:block}
\end{figure}

 The SiPM signal is amplified and sent to a comparator and a peak detector circuit. The comparator triggers data acquisition by providing a timestamp, while the peak detector holds a constant fraction of the pulse peak for accurate amplitude measurement by the Raspberry Pi Pico (RPi) MCU~\cite{raspberrypi_pico}. The resulting ADC value is proportional to the number of incident photons. Additionally, the SiPM signal is available through a BNC connector for external readout with nanosecond timing resolution (see Sec.\ref{sec:velocity}). A block diagram of the main board is illustration in Fig.~\ref{fig:block}.

To suppress the radiogenic background signals and focus on cosmic-ray muons, the v3X detector supports coincidence detection using multiple units. In a two-detector configuration, each unit independently processes events as previously described; however, a coincidence event, defined as a trigger in both detectors within a predefined time window, is flagged when both comparators fire nearly simultaneously. This temporal correlation effectively reject uncorrelated background events, such as environmental gamma and beta radiation, which are unlikely to occur in both spatially separated detectors at the same time. Coincidence detection enables the identification of through-going cosmic-ray muons and forms the basis for more advanced applications, including muon tracking geometries and energy deposition studies (see Sec.~\ref{sec:full_sky}), and altitude measurements (see Sec.~\ref{sec:hab}).

Once triggered, the RPi records the event count, timestamp, 12-bit pulse amplitude, SiPM pulse height, temperature, barometric pressure, linear acceleration, angular velocity, cumulative dead time, and whether or not the event was seen by a coincident detector. The master and coincidence count rates are shown on an OLED display, while every event’s data is recorded to a microSD card and streamed in real time via USB port. If data is saved directly to a computer using the provided software, an additional time and date stamp are provided through Network Time Protocol (NTP). This enables global synchronized real-time information and removes the internal drift of the MCU oscillator. All data is saved as easy to read, tab-delimited .txt file, with each row corresponding to an individual triggering event. Once an event has been fully recorded, a metal-oxide-semiconductor field-effect transistor (MOSFET) shunts the peak detector charging capacitor to ground, initiating the search for subsequent pulses.

Although users do not have direct access to the firmware, when a microSD card is inserted, a file named configure.txt is created, which contains variables that can be changed to provide some level of flexibility. This includes options to change the detector trigger threshold, rename the detector, turn off the OLED screen and external LED lights, 
and enable/disable the buzzer. 

The detector itself can be powered through a  USB port \color{black}on a computer, using a small battery, power bank, a mobile phone, or through the coincidence cable. A single detector consumes 0.5\,W. For context, a commonly available 5,000\,mAh power bank was measured to power two detectors for 18 hours. The total mass of the detector, including the optional aluminium case, is 110\,g (83\,g without) and measures 66.4\,mm$\times$100.0\,mm$\times$39.9\,mm (63.2\,mm$\times$100.0\,mm$\times$31\,mm without). 


By connecting two v3X detectors using a standard CAT5/6 cable via the coincidence port (see figure~\ref{fig:cosmicwatch}), users can enable \textit{coincidence detection}, enabling accurate identification of cosmic-ray muons and selection of the incoming particle’s direction. The muon rate at the sea-level is approximately 1$\mathrm{cm^{-2}min^{-1}}$~\cite{PDG_CosmicRays}, which corresponds to  0.42\,Hz given the 25\,cm$^2$ scintillator employed in the v3X detector.  




\color{black}
\section{Detector Component-Level Design}

\subsection{Scintillator and single photon sensor}
The detector employs a 5\,$\times$\,5\,$\times$\,1\,cm$^3$ slab of organic plastic scintillator, composed of a polystyrene base doped with 1\% by weight of PPO (2,5-diphenyloxazole) and 0.03\% POPOP (1,4-bis[2-(5-phenyloxazolyl)]benzene)~\cite{beznosko2004fnal}. This composition yields approximately 10,000\,photons/MeV of deposited energy. The scintillator exhibits a maximum emission at 420\,nm (violet) and an absorption cut-off below 400\,nm, matching well the SiPM spectral sensitivity. The material is similar in formulation to the FNAL-NICADD extruded scintillator studied for large-scale calorimeters such as ALICE EMCal~\cite{grachov2005study} and the MINOS calorimeter~\cite{pla2001extruded}.


\begin{figure}[b]
    \centering
    \includegraphics[width=12cm]{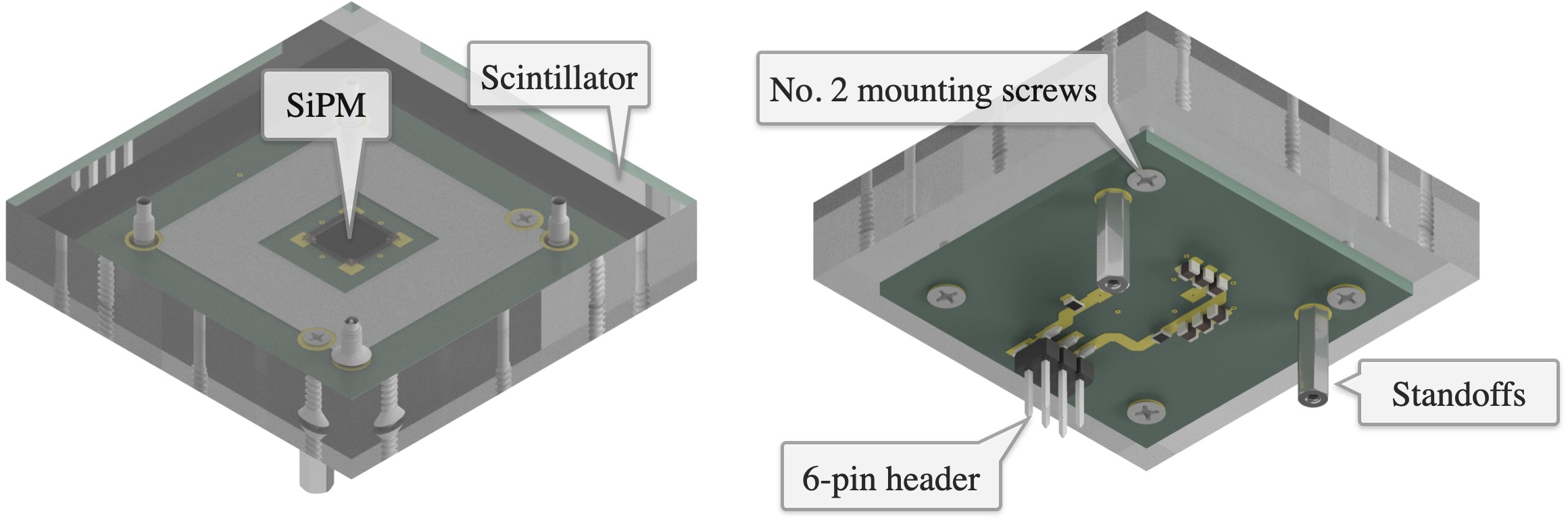}
    \caption{A rendering of the top  and bottom  sides of the SiPM PCB mounted to the plastic scintillator with four 3/8" \#2 screws. }
    \label{fig:sipmPCB}
\end{figure}

 Scintillation photons are either internally reflected at the scintillator–air interface or, if they escape, redirected by a surrounding layer of reflective aluminum foil. A portion of these photons reach a photosensor, an OnSemiconductor MicroFC (60035 C-Series) SiPM, \color{black}which measures 7\,mm$\times$7\,mm, with a 6\,mm$\times$6\,mm active area~\cite{SiPM}. The SiPM is mounted on a custom 2-layer PCB and operated at 30\,V reverse bias. This corresponds to an overvoltage (voltage above the breakdown voltage) of 5.5\,V, achieving a photon detection efficiency of about 43\% at 420\,nm.

To optimize light collection, the SiPM is optically coupled to the scintillator using PMT optical gel and/or a 0.3\,mm-thick silicone pad. This coupling minimizes reflection losses at the interface by more closely matching the refractive indices of the scintillator and the SiPM's acrylic window. The scintillator and SiPM PCB are wrapped in aluminum foil for reflectivity, and then optically isolated using three to four orthogonal layers of black electrical tape.

The SiPM-scintillator assembly connects to the main PCB via a 6-pin connector and is mechanically secured with two 7/16"-long 0-80 aluminum hex standoffs. A rendering of the SiPM PCB, excluding the aluminum foil and optical isolation layers, is shown in figure~\ref{fig:sipmPCB}.


\subsection{The main PCB}
\begin{figure}[b]
    \centering
    \includegraphics[width=\textwidth]{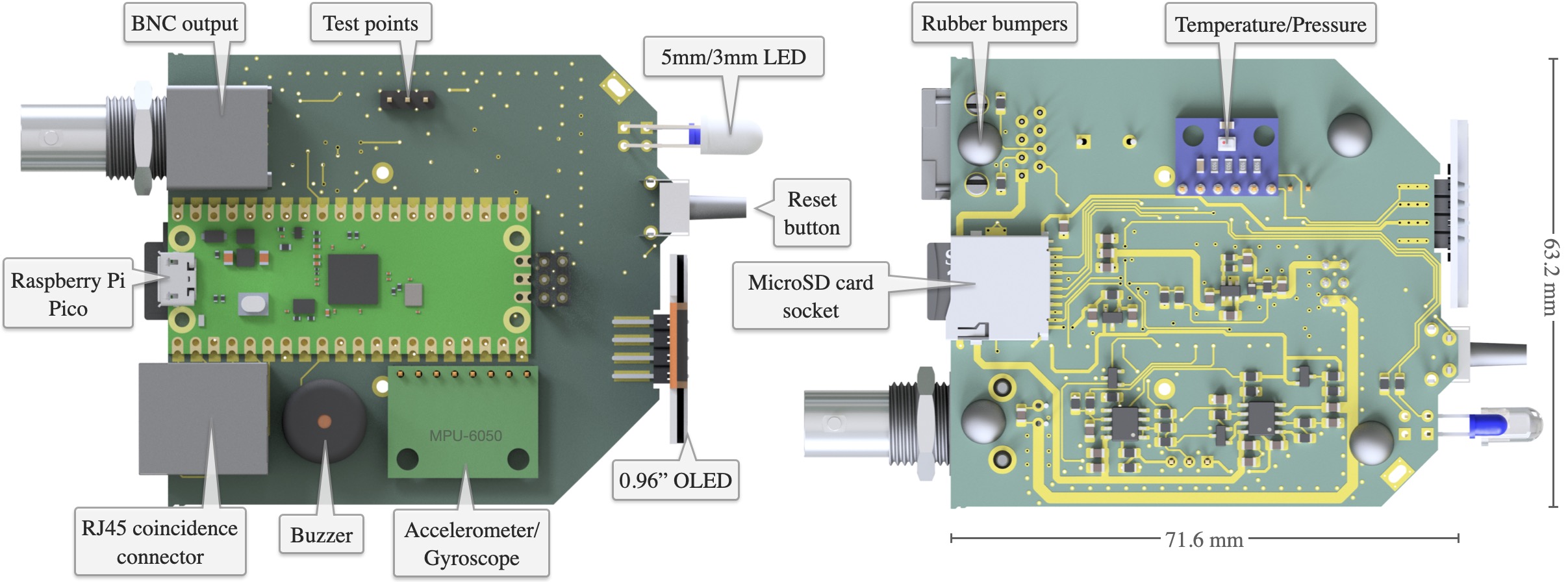}
    \caption{Rendering of the top and bottom of the main PCB.}
    \label{fig:PCB}
\end{figure}

The main PCB supplies power and bias voltage to the detector, hosts the analog pulse-shaping circuitry, and serves as the backbone for the digital peripherals and the RPi. As shown in figure~\ref{fig:PCB}, the top of the board hosts the majority of the digital components (with the exception of the temperature \& pressure sensor and microSD card socket), while the bottom of the board contains the powering systems and analog electronics. The full electronic schematic for the CosmicWatch v3X is presented in figure~\ref{fig:schematic} of Appendix A.

\subsubsection{Powering systems}
All on-board electronics are powered via the 5\,V USB connector on the RPi. After passing through a protection diode, the system voltage of 4.5\,V supplies the DC-DC booster for the SiPM and powers the two dual operational amplifiers. A regulated 3.3\,V line from a regulator on the MCU powers the digital peripherals and feeds a 2.5\,V precision reference voltage used as a reference for the ADC and biasing the inputs to the operational amplifiers. The 5\,V line also feeds through the RJ45 coincidence connector, allowing a single USB source to power multiple daisy-chained detectors.



The SiPM requires a low-ripple, stable bias of greater than $24.5\pm0.2$\,V. An over-voltage can be supplied to enhance the gain of the SiPM and photon detection efficiency. The v3X detector employs a MAX5026 DC–DC booster to convert the system voltage to 30.0\,V, corresponding to an over-voltage of 5.5\,V. The output of the DC-DC booster is passed through a RC filter and local decoupling capacitor at the SiPM 6-Pin header. Through optimized component layout and component selection the measured DC-DC booster ripple is found to be $\le$10\,mV.

\subsubsection{Analog pulse shaping}

The raw SiPM pulse is first routed to a back-panel BNC connector for high-fidelity signal readout with nanosecond-level timing resolution. Simultaneously, it is sent to the input of an AC-coupled TPH2502 operational amplifier, configured in a non-inverting mode with a gain of 22.6\,dB (voltage gain of 13.5\,V/V) and a bandwidth of approximately 10\,MHz. A 25\,mV DC offset, derived from a voltage divider tied to the 2.5\,V reference, shifts the AC coupled signal away from the operational amplifier's negative rail. A high-pass filter (cutoff $\sim$3\,kHz) removes low-frequency drift. After amplification, the signal is re-biased to 25\,mV and sent to both the peak detector and comparator circuits. The amplified signal can be seen by connecting an oscilloscope to Test Point One (TP1).



The peak detector circuit employs a Schottky diode to charge a small capacitor during each incoming pulse. This voltage is then buffered by a secondary operational amplifier and scaled to the 0–2.5\,V ADC input range using a resistive voltage divider. The output of the peak detector can be monitored at Test Point Two (TP2) using a high-impedance oscilloscope input and a short coaxial cable to minimize reflections. The ADC samples this voltage to estimate the amplitude of the original SiPM pulse. After digitization, a MOSFET controlled by the RPi discharges the capacitor, resetting the peak detector for the next event.

The comparator uses another TPH2502 operational amplifier to compare the signal to a threshold voltage set by a RC filtered pulse width modulation (PWM) output from the RPi, providing a tunable DC reference voltage. If the pulse exceeds the threshold, the comparator outputs a 4.5\,V logic pulse whose duration matches the time over threshold of the amplified signal. This pulse is scaled down to 3.3\,V through a voltage divider for compatibility with the RPi. The trigger threshold can be adjusted in the configure.txt file located on the microSD card. 

Together, the pulse shaping and triggering system allows the detector to determine the SiPM pulse height and the time of the each event with high precision and low noise.

\subsubsection{Digital electronics}

At the heart of the detector's digital system is the RPi, a dual-core MCU operating at 133\,MHz. One core handles time-critical data acquisition tasks, such as reading the ADC, recording timestamps, measuring the dead time, resetting the peak detector, and printing the event data to the serial port, while the second manages peripheral communication, like the OLED display, microSD card, and environmental sensors. This separation significantly reduces dead time (see Table~\ref{tab:specifications}) and ensures responsive operation even at hundreds of events per second. 

The RPi interfaces with several digital peripherals via I\textsuperscript{2}C and SPI buses, including:
\begin{itemize}
    \item A Bosch BMP280 sensor for temperature and barometric pressure  monitoring;
    \item A microSD card for local data logging and configuration;
    \item An OLED display for real-time readout of the measurement and system status;
    \item An onboard buzzer for audible feedback;
    \item A 6-axis inertial measurement unit (IMU) accelerometer and gyroscope to monitor the orientation and motion tracking.
\end{itemize}

The system supports USB serial communication (at a baud rate of 115,200\,bps), enabling real-time monitoring and firmware updates. When data is captured via USB port on a computer using the provided import\_data.py software, each event is timestamped in real time by the host computer. These timestamps typically synchronize to the NTP, ensuring accuracy within hundreds of microseconds to a few milliseconds.

The coincidence port, implemented as an RJ45 (8P8C) connector, allows two detectors to communicate with each other. When two detectors are connected through a CAT5 or CAT6 cable and rebooted, they automatically enter \textit{coincidence mode}, indicated by flashing the blue and white LEDs for one second. In this mode, triggers from each detector are exchanged, and events that occur within approximately 2.3\,$\mu$s of each other are labeled as coincident by both detectors. Additionally, the coincidence port provides future expandability through available I$^2$C and SPI interfaces. These will eventually facilitate communication with external peripherals, such as GPS modules, magnetometers, real-time clocks, and other sensors.

Collectively, these digital electronics create a flexible, low-latency platform suited for detector control, efficient data acquisition, and straightforward integration with external peripherals. These detectors are designed to operate reliably within a temperature range of approximately –20 °C to +85 °C.

\subsection{Detector assembly}
A full instruction manual is available on the Github repository\footnote{\url{https://github.com/spenceraxani/CosmicWatch-Desktop-Muon-Detector-v3X}}, which contains the information needed to purchase the components and build a complete detector. It amounts to purchasing the list of components, and following guided instructions on manufacturing the PCBs and populating each part of the circuit sequentially. 

\begin{figure}[h]
    \centering
    \includegraphics[width=\linewidth, angle=0]{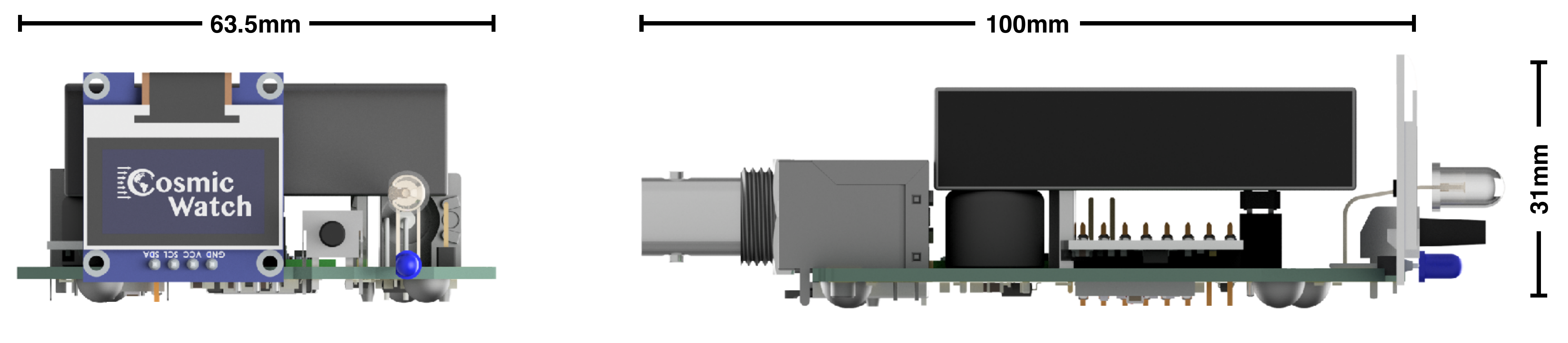}
    \caption{The SiPM and main PCB assembly.}
    \label{fig:internal}
\end{figure}

\begin{figure}[h]
    \centering
    \includegraphics[width=\linewidth, angle=0]{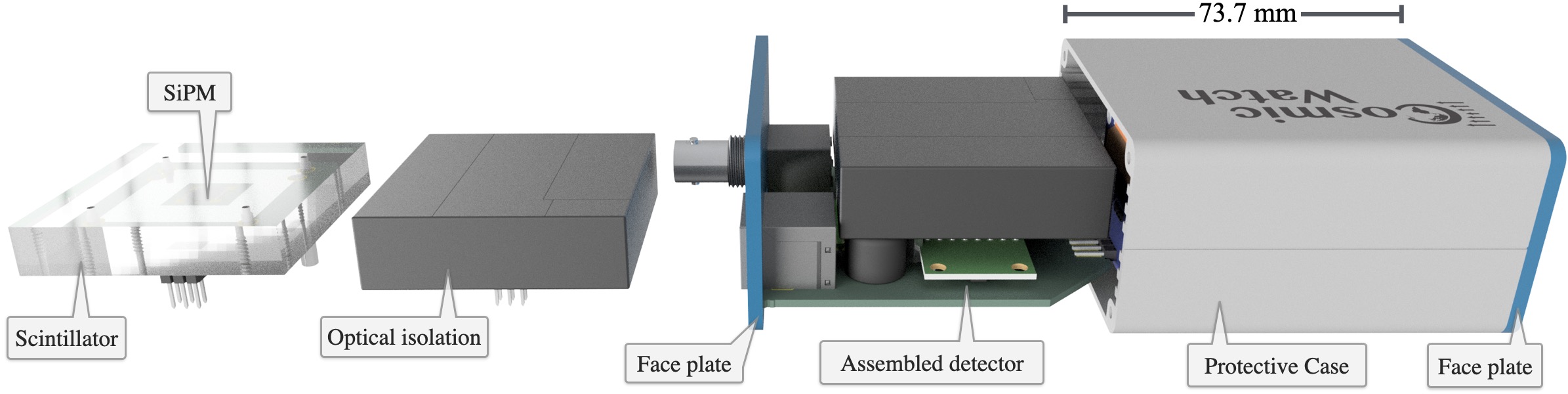}
    \caption{The full assembly of the CosmicWatch v3X detector.}
    \label{fig:assembly}
\end{figure}

Once the SiPM and main PCBs are assembled (figure~\ref{fig:internal}), the 6-pin header on the SiPM board inserts into the corresponding 6-pin socket on the main PCB and is secured using two hex standoffs. The full assembly then slides into the rails of the aluminum enclosure and is held in place by two laser-cut faceplates, as shown in figure~\ref{fig:assembly}.

Uploading the firmware to operate the detector is straightforward. The user may simply hold down the BOOTSEL button on the RPi while connecting to a computer using a micro USB cable (with data capabilities). The device will appear as a removable storage device. The firmware, provided as a .uf2 file, can be simply dragged and dropped onto the device, and the detector will automatically reboot with the new firmware installed.

\section{Analog Signal Processing Calibration}\label{sec:calibration}
\begin{figure}[http!]
    \centering
    \includegraphics[width=\textwidth,angle=0]{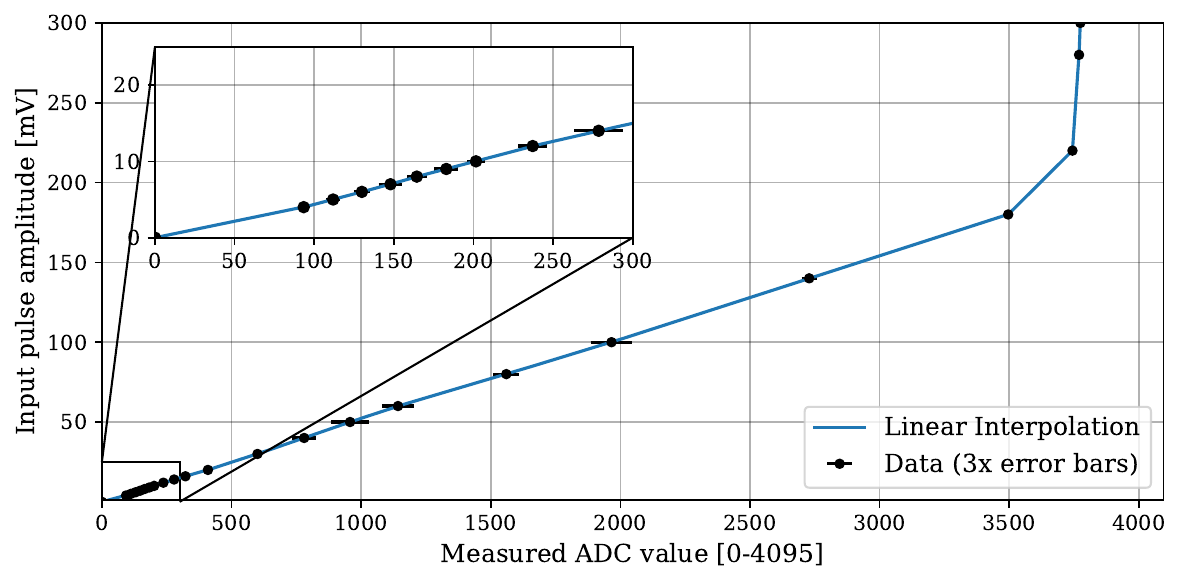}
    \caption{The calibration curve (blue) which converts the measured peak detector ADC values into SiPM pulse-height equivalents. At a input pulse amplitude of approximately 180\,mV the electronics begin to saturate.}
    \label{fig:calibration}
\end{figure}

After measuring the signal amplitude from the peak detector, the next step is to convert this value to the corresponding SiPM peak voltage, which is approximately proportional to the energy deposited in the scintillator. This calibration is performed by temporarily removing the SiPM and scintillator assembly and injecting test pulses of known amplitude into the system via the BNC input. By recording the peak detector’s ADC output for a range of known input pulse amplitudes, a relationship between the measured ADC values and the actual SiPM peak voltage can be established. This procedure is conducted over a series of input levels using an arbitrary waveform generator (AWG), which delivers 100 identical pulses for each amplitude setting. The resulting data, shown in figure~\ref{fig:calibration}, is used to linearly interpolate a calibration curve, allowing conversion of measured peak detector ADC values into SiPM peak voltage equivalents.

\section{Firmware Features and Functionality}
The CosmicWatch v3X firmware is written in C++ and runs on the RPi MCU, which features dual ARM Cortex-M0+ cores. The code-base is organized into modular components, each responsible for managing a specific subsystem of the detector. These modules interface with the microSD card, OLED display, environmental sensors (BMP280 and MPU-6050), and core utility functions. Upon startup, the firmware initializes all subsystems and performs a handshake over the coincidence pins to detect the presence of a second detector. If a positive response is received, \textit{coincidence mode} is activated, and the user is notified with a bright one-second flash of the LEDs.

Once initialization is complete, the system transitions to continuous operation. The firmware takes advantage of the RPi's dual-core architecture: Core\,0 is exclusively responsible for real-time data acquisition, while Core 1 performs non-time critical auxiliary tasks such as display updates and microSD card data logging.

On Core\,0, a high-speed digital polling loop continuously monitors the output of a comparator. Upon detecting a rising edge, indicating a particle interaction in the scintillator, the firmware initiates a 2.3,$\mu$s coincidence window, during which the second coincidence pin is polled 32 times. If a HIGH signal is observed in any of these polls, the event is flagged as coincident. The firmware then samples the peak ADC value from the waveform and records a timestamp, as illustrated in Fig. 8. Visual feedback is provided via onboard LEDs: white for all events and blue for coincident events.

\begin{figure}[t]
    \centering
    \includegraphics[width=\textwidth]{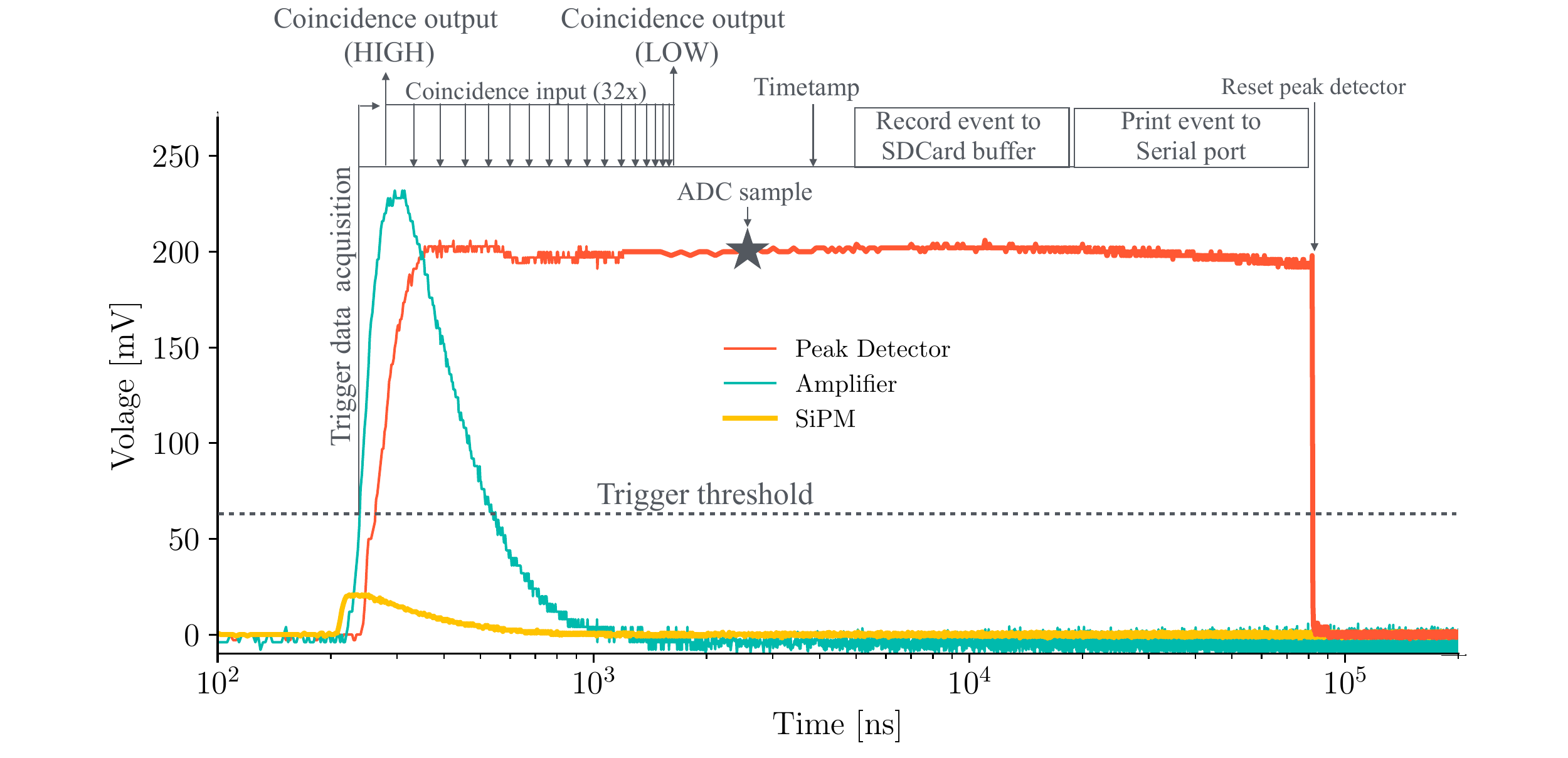}
    \caption{Logic diagram of the CosmicWatch data acquisition chain. Signals from the SiPM coupled to the scintillator are amplified and passed through a peak detector. The trigger logic compares the signal to a threshold, records coincidences, and manages reset. Events are timestamped and written to the SD card buffer or printed to the serial port. }
    \label{fig:logic}
\end{figure}

A data packet is constructed containing a timestamp, the peak signal measurement, and values from all available environmental sensors. This packet is transmitted over the USB serial port and simultaneously added to a primary ring buffer capable of holding 4096 events. To avoid buffer overflow during high event rates, artificial dead time is introduced when the buffer nears 80\% capacity.

Core\,1 runs an independent loop that handles user-facing tasks and data logging. It periodically updates the OLED display and collects environmental data including barometric pressure, temperature, acceleration, and angular velocity. During each cycle, Core\,1 checks the primary buffer for new events and transfers them to a secondary 32\,kB circular buffer, capable of storing approximately 250 formatted events. This intermediate buffer is flushed to the microSD card in batch mode either when it exceeds a predefined threshold (20 events) or after a 10-second interval.

This dual-buffer architecture minimizes write latency, enables sustained event logging rates of at least 700\,Hz to both the serial port and microSD card, and significantly extends the microSD card’s lifespan by reducing the frequency of write operations. For applications requiring higher throughput, a configurable option in the config.txt file allows users to disable serial communication, increasing the maximum logging rate to the microSD card to approximately 7,000\,Hz.

Additional safety and debugging features are built into the firmware. On startup, the SiPM bias voltage and trigger threshold are measured via ADC channels 1 and 2 to ensure they fall within specified tolerances. All critical system parameters including, startup diagnostics and event records are streamed in real time through the USB serial interface. The system is robust against data loss and can operate with or without a microSD card, defaulting to serial only output when the card is not present.

\section{Experimental Setup and Data Collection}
\subsection{Detector orientation}

\begin{figure}[h]
\begin{center}
\includegraphics[width=1.0\columnwidth]{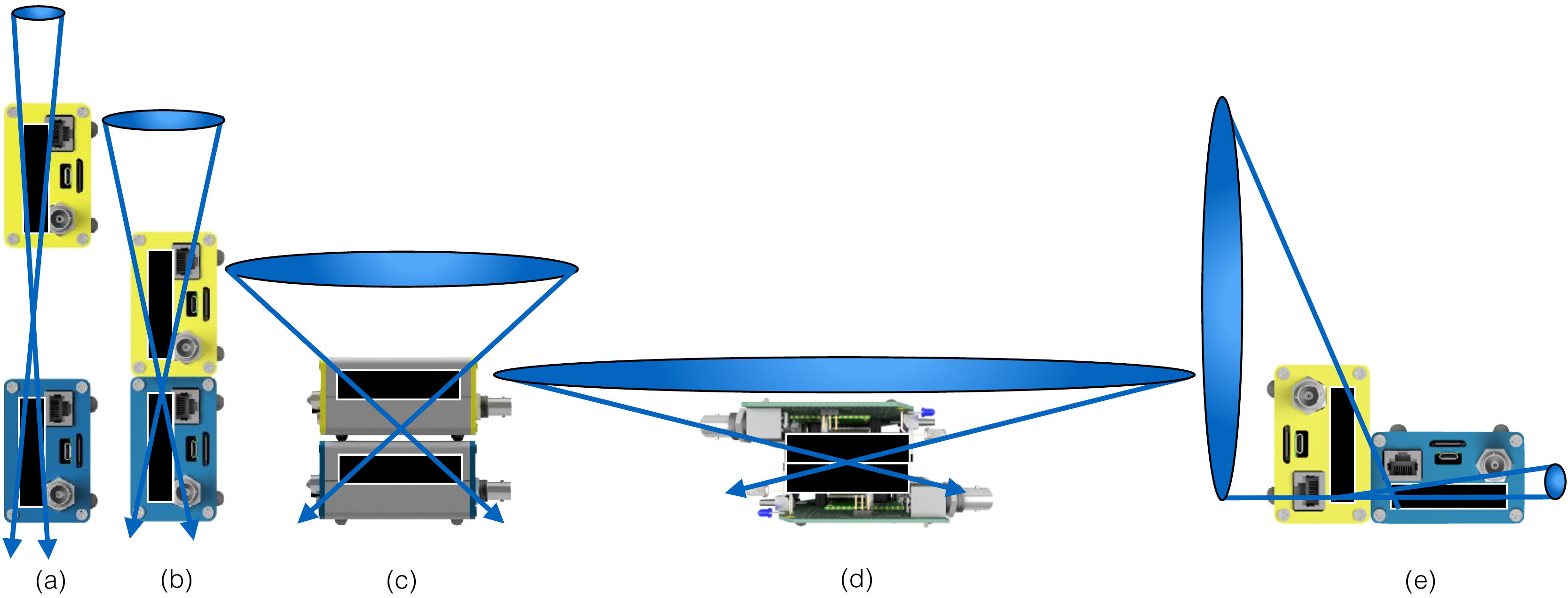}
\caption{Several configurations for setting up the detectors in \textit{coincidence mode}. Moving from left (a) to right (d) the solid angle subtended between detectors increases (illustrated as the blue areas above the detectors) as well as the coincidence rate as measured by the coincident detector. Directionality can be inferred with configuration (e), which assumes cosmic-ray muons are not traveling upwards.} 
  \label{fig::opening_angle}
 \end{center}
\end{figure}

When conducting a coincidence measurement between two detectors, careful consideration of what is being measured and how the orientation impacts the measurement is important. For instance, when interested in the angular spectrum of cosmic-ray muons, the goal is to selectively accept muons originating from a small solid angle (a specific area of the sky). figure~\ref{fig::opening_angle} illustrates several possible configurations. On the left side of figure~\ref{fig::opening_angle}, labeled as (a), two detectors spaced a few centimeters apart are connected with a coincidence cable. In this setup, only muons traveling downward through the solid angle represented by the blue  area can trigger both detectors, a practical measurement using this configuration is described in Sec.~\ref{sec:angle}. Figure~\ref{fig::opening_angle} (d) depicts a configuration where both detectors will trigger from down-going muons over a much larger solid angle. This configuration is commonly employed to extract a full sky muon rate, as in Sec.~\ref{sec:full_sky}

\subsection{Data acquisition methods}
There are several ways to save data from the detector. The OLED provides realtime rate measurements, while the microSD card and USB serial connection can be used to see detailed event information. Each recorded event is approximately 85\,bytes in size.  

\subsubsection{OLED readout}
The OLED display is ideal for users who only need overall event counts and average event rates, without detailed per‑event data. It reports the total run time, number of triggered events, and the average count rate (accounting for dead time). If the detector boots into coincidence mode, two additional rows indicating the number of coincident events and coincidence average count rate are also reported. Additionally, the OLED display reports the microSD card file name (if present), and the detector name (modifiable in the configure.txt file located on the microSD card). The OLED can also be turned on/off using the configure.txt file.

\subsubsection{MicroSD card}
\begin{figure}[http!]
\centering
\includegraphics[width=15cm]{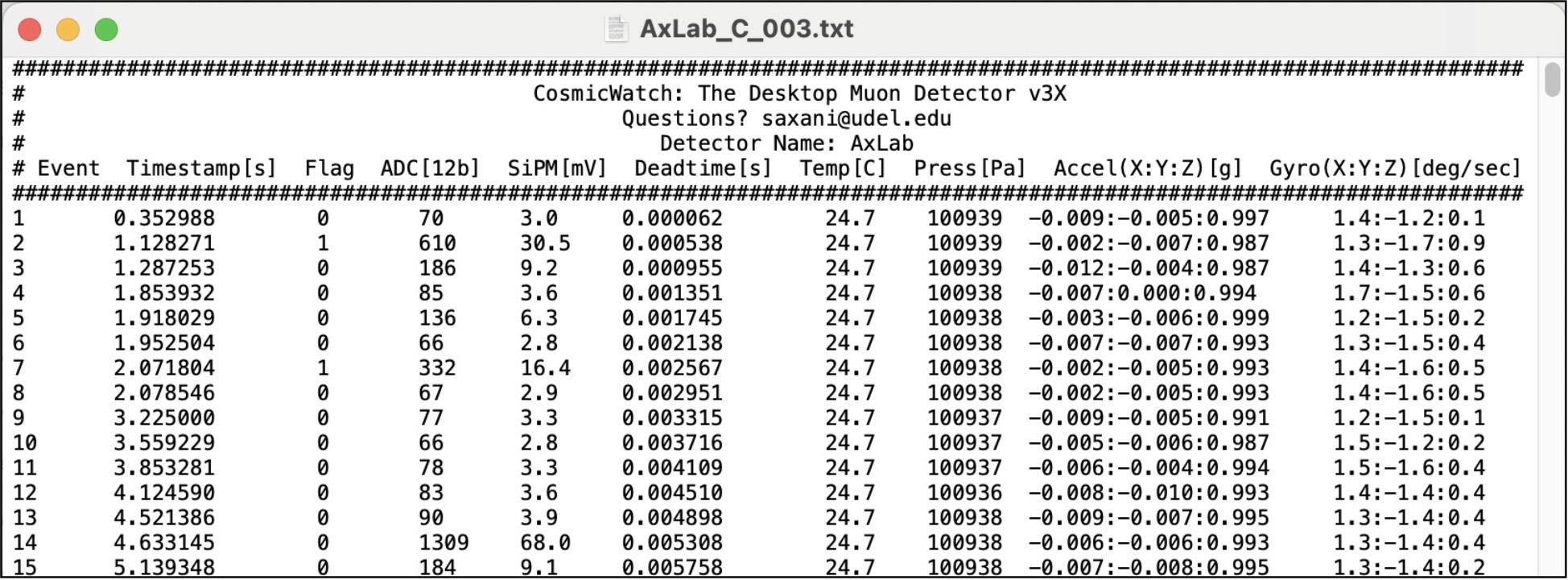}
\caption{Example data file from the microSD card. Events 2 and 7 are flagged as coincident events.}
\label{fig:sd_data}
\end{figure}

If a microSD card is inserted, the detector creates a new file each time it is powered on or reset. The microSD card must be formatted as exFAT. MicroSD card filenames begin with the detector name (configurable to fewer than 20 characters via the configure.txt file); followed by a mode indicator: "C" if the detector booted in \textit{coincidence mode}, or "M" if in \textit{master mode}; and a number that increments sequential upwards from the last saved file. Data is stored in plain text format with a .txt extension, as illustrated in the coincident data from a detector named "AxLab" operating in \textit{coincidence mode} in figure~\ref{fig:sd_data}. A relatively small microSD card ($\ge$4\,GB) enables users to save several months worth of continuous data at sea level.

The data shown in the figure~\ref{fig:sd_data} is formatted into tab-delimited columns. Each column is labeled in the header and defined as follows:

\begin{itemize}
    \item \textbf{Event}: The event number. Sequentially counts upwards on each trigger.
    \item \textbf{Timestamp [s]}: The total run time (not live time), measured in seconds. This is only accurate to roughly tens of seconds per day given the uncertainty by the precision of the RPi internal clock. 
    \item \textbf{Flag}: \textit{Coincidence mode} flag. If the detector observes a coincidence signal while operating in \textit{coincidence mode}, the event is flagged with a boolean "1." Non-coincident events are flagged with a "0." 
    \item \textbf{ADC [12b]}: The 12-bit (0-4095) ADC measurement of the buffered peak detector output. The ADC is referenced to 2.5\,V.
    \item \textbf{SiPM [mV]}: The calculated SiPM pulse height is based on the ADC measurement. This value is proportional to the energy deposited in the scintillator (see Sec.~\ref{sec:calibration}). 
    \item \textbf{Dead time [s]}: The cumulative dead time since the detector start. Dead time should be accounted for when making any rate measurement. To calculate the event rate, divide the total counts by the difference between the total run time and the total dead time. The denominator is what is called live time, and represents the amount of time the detector was actively searching for events.
    \item \textbf{Temp. [$^\circ$C]}: The local temperature, in degrees Celcius, measured by the BMP280 sensor.
    \item \textbf{Press. [Pa]}: The barometric pressure, in pascals, measured by the BMP280 sensor.
    \item \textbf{Accel. (X:Y:Z) [g]}: The x-y-z acceleration measured in units of gravitational acceleration \textit{g} ($g \approx 9.81$m/s$^2$). 
    \item \textbf{Gyro. (X:Y:Z) [deg/s]}: The x-y-z angular acceleration in degrees per second. 
\end{itemize}

\subsubsection{USB connection to a computer}
There are three ways to take data in real time with a computer, capturing data from a USB port at a BAUD rate of 115,200\,bps.

\paragraph{The import\_data.py script:} Running import\_data.py (provided in the Data directory on GitHub) from the host computer prompts selection of the USB port corresponding to the connected detector, followed by specification of a file name for storing the incoming data. Data acquired with the import\_data.py script include three additional columns beyond the standard output.

\begin{itemize}
    \item \textbf{Detector Name}: This column displays the detector name. It is useful if the user selects to save the data from multiple detectors to a single file. 
    \item \textbf{Time stamp}: An additional realtime time stamp from the computer, taken from the computer. This is from NTP and has millisecond precision. 
    \item \textbf{Date}: The date of the event, taken by the computer.
\end{itemize}


Additionally, the script provides the capability to connect to the CosmicWatch website, which can be used to upload or record data, and visualize the results in real time. 

\paragraph{A Serial monitor:} Serial monitors are available that allow the data to be viewed in real time. When the detector is connected to a host computer via a Micro-USB cable that supports both power and data transfer. Data are transmitted asynchronously at a baud rate of 115,200 bps. A convenient option is to use Visual Studio Code (VS Code) with the “Serial Monitor” extension installed. The Serial Monitor is particularly useful for troubleshooting, as the detector outputs a set of diagnostic messages during bootup that can indicate potential problems.

\begin{figure}[http!]
\centering
\includegraphics[width=\textwidth]{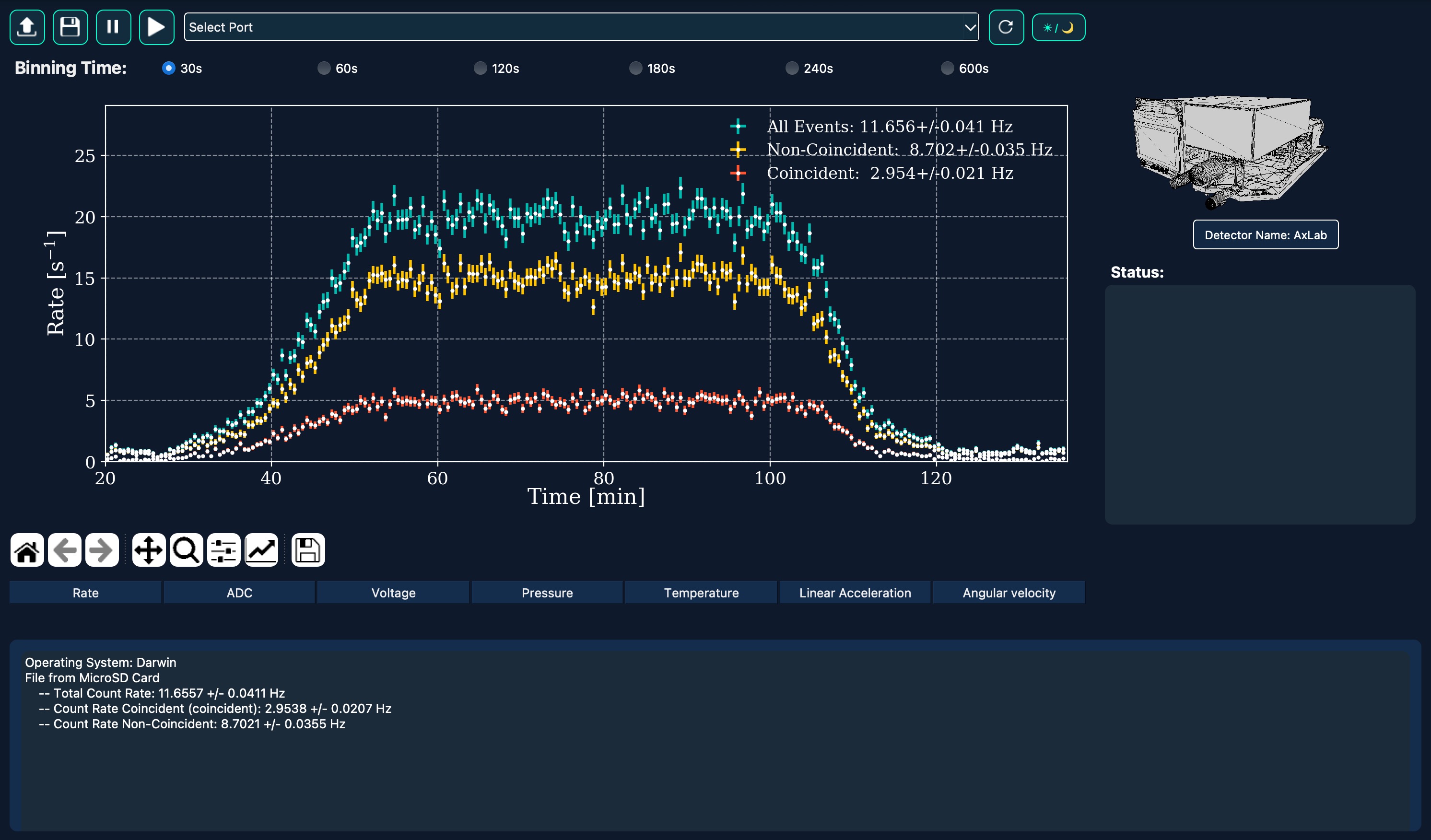}
\caption{Graphical User Interface (GUI) of the CosmicWatch data acquisition software. The interface enables both real-time data recording to the computer and offline plotting of data from previously saved files. It provides a straightforward way to visualize various distributions, binned over user-selectable time intervals. The example shown here uses data collected with two detectors in configuration (c) of figure~\ref{fig::opening_angle}, during a flight from Philadelphia (PHL) to Madison, WI, binned in 20s intervals.}
\label{fig:gui}
\end{figure}

\paragraph{The Graphical User Interface (GUI):} The CosmicWatch graphical user interface (GUI; figure~\ref{fig:gui}) offers an accessible platform for data visualization, either in real time through a serial port connection or in post-processing using uploaded \texttt{.txt} files. Users can explore coincident and non-coincident event rates, ADC distributions, SiPM pulse-height spectra, as well as pressure, temperature, linear acceleration, and angular velocity measurements. The GUI further provides flexible binning options, live count displays, and can function as a serial monitor to record data directly from the selected port. Its intuitive layout facilitates efficient evaluation of detector performance and the production of publication-ready figures.




\newpage
\section{Example measurements}\label{sec:measurements}

\subsection{A full-sky cosmic-ray muon measurement}
\label{sec:full_sky}

Two v3X detectors were arranged in configuration (d), as shown in figure~\ref{fig::opening_angle}, in the basement of Sharp Lab at UD to maximize the solid-angle acceptance to $\Omega \approx 2\pi$\,sr. Data presented in figure~\ref{fig:sd_data} were recorded continuously to a microSD card over 26.9\,h, yielding a total of $N_{\rm total} = 2.35 \times 10^5$ events and $N_{\rm coin} = 3.05 \times 10^4$ coincident events. These correspond to a total trigger rate of $R_{\rm total} = 2.423 \pm 0.005$\,Hz and a coincidence rate of $R_{\rm coin} = 0.315 \pm 0.002$\,Hz, or $0.756 \pm 0.005$\,cm$^{-2}$min$^{-1}$. The total dead time during this measurement was 0.1\% of the live time, corresponding to an average dead time per event of 408\,$\mu$s.




\begin{figure}[b]
    \centering
    \begin{subfigure}[b]{0.481\textwidth}
        \centering
        \includegraphics[width=\textwidth]{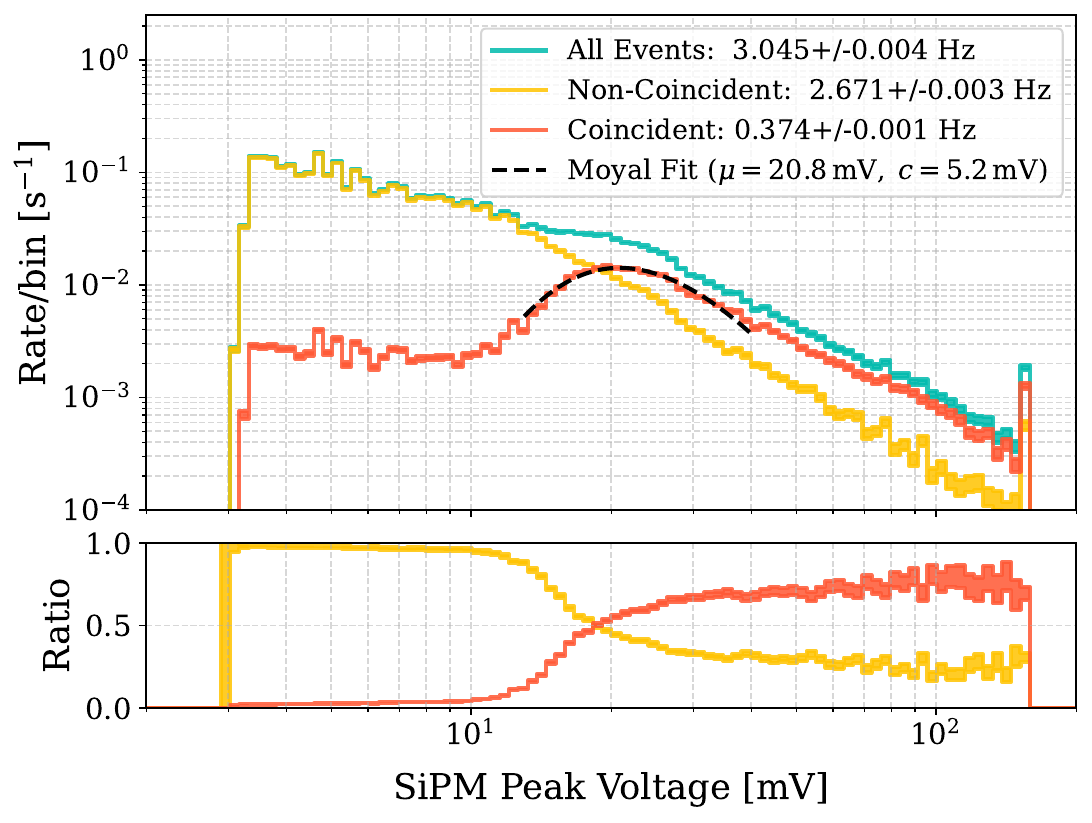}
        \label{fig:spectrum1}
    \end{subfigure}
    \begin{subfigure}[b]{0.51\textwidth}
        \centering
        \includegraphics[width=\textwidth]{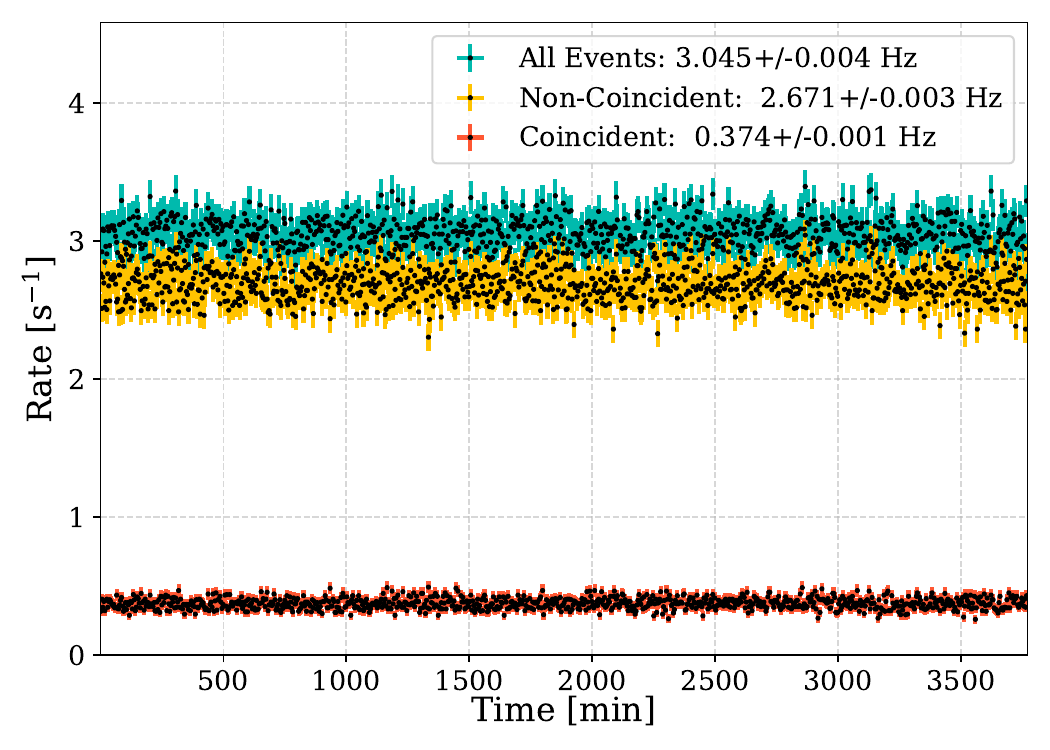}
        \label{fig:spectrum2}
    \end{subfigure}
    \caption{Left: the SiPM pulse‐height spectra recorded by a set of detector in configuration (d). The coincident events begin dominating at SiPM pulse-heights greater than approximately 18\,mV. Right: The measured detection rate of coincident and non-coincident events. }
    \label{fig:mip_spectra}
\end{figure}

The SiPM pulse‐height spectrum (figure~\ref{fig:mip_spectra}, left), converted from ADC counts using the calibration curve in figure~\ref{fig:calibration}, exhibits the characteristic Landau‐shaped “minimum ionizing particle (MIP) bump” produced by through‐going muons in a thin absorber. The Landau energy‐loss distribution can be approximated by the analytic Moyal distribution~\cite{kolanoski2020particle,moyal1955ionization}:
\begin{equation}
f(x;\mu,c)
= \frac{1}{\sqrt{2\pi}\,c}
  \exp\!\Bigl[-\tfrac12\bigl(z + e^{-z}\bigr)\Bigr]
\quad\text{with}\quad
z = \frac{x - \mu}{c}\,,
\end{equation}
where \(\mu\) denotes the most probable value and \(c\) the width parameter. A fit over the range \(13\,\mathrm{mV}\) to \(40\,\mathrm{mV}\) yields \(\mu \approx 21\,\mathrm{mV}\), which corresponds to the \(\sim2\,\mathrm{MeV}\) energy loss of a minimum‐ionizing muon traversing the \(\approx1\,\mathrm{cm}\)-thick plastic scintillator~\cite{groom2000passage}, and \(c \approx 5\,\mathrm{mV}\). This is only approximate, as the average trajectory of the muons through the scintillator will be slightly larger than \(1\,\mathrm{cm}\). 
Detector saturation appears near \(180\,\mathrm{mV}\), as denoted by the overflow bin. The relatively flat tail below \(10\,\mathrm{mV}\) in the coincident‐event distribution is attributed to corner‐clipping muons, which graze the scintillator edge and deposit reduced energy.

This demonstrates the ability to select muons using \emph{coincidence mode} and shows how the SiPM pulse amplitude can be used to further discriminate between muons and radiogenic backgrounds based on energy deposition. The ratio plotted in the left subplot of figure~\ref{fig:mip_spectra} indicates that above approximately \(18\,\mathrm{mV}\), cosmic‑ray muons dominate the energy deposition spectrum.

The precise location of the MIP peak depends on the precise value of the breakdown voltage as well as the quality of the detector assembly: the efficiency of the optical coupling between the SiPM and scintillator, the scintillator surface finish, the reflectivity of any wrapping, and other factors that influence photon collection. It also shifts with the SiPM bias voltage, which is set by precision (0.1\,\%) resistors and directly determines the gain. Nevertheless, because the MIP peak occurs well above the trigger threshold ($\approx$4\,mV), even substantial variations in collection efficiency have minimal effect on muon detection efficiency; only corner‑clipping events register near threshold. Overall count rates, however, are expected to show more significant detector to detector variations due to the steep rise of the non‑coincident background near threshold.

\begin{figure}[ht]
    \centering
    \begin{subfigure}[b]{0.49\textwidth}
        \centering
        \includegraphics[width=\textwidth]{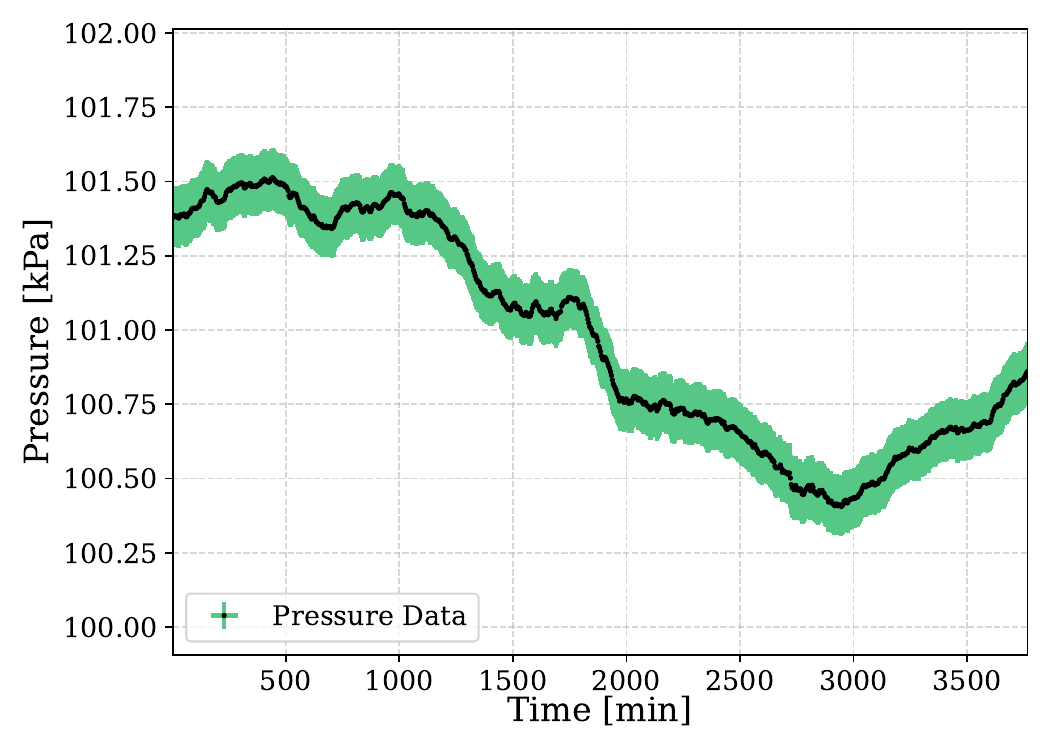}
        \label{fig:pressure}
    \end{subfigure}
    \begin{subfigure}[b]{0.49\textwidth}
        \centering
        \includegraphics[width=\textwidth]{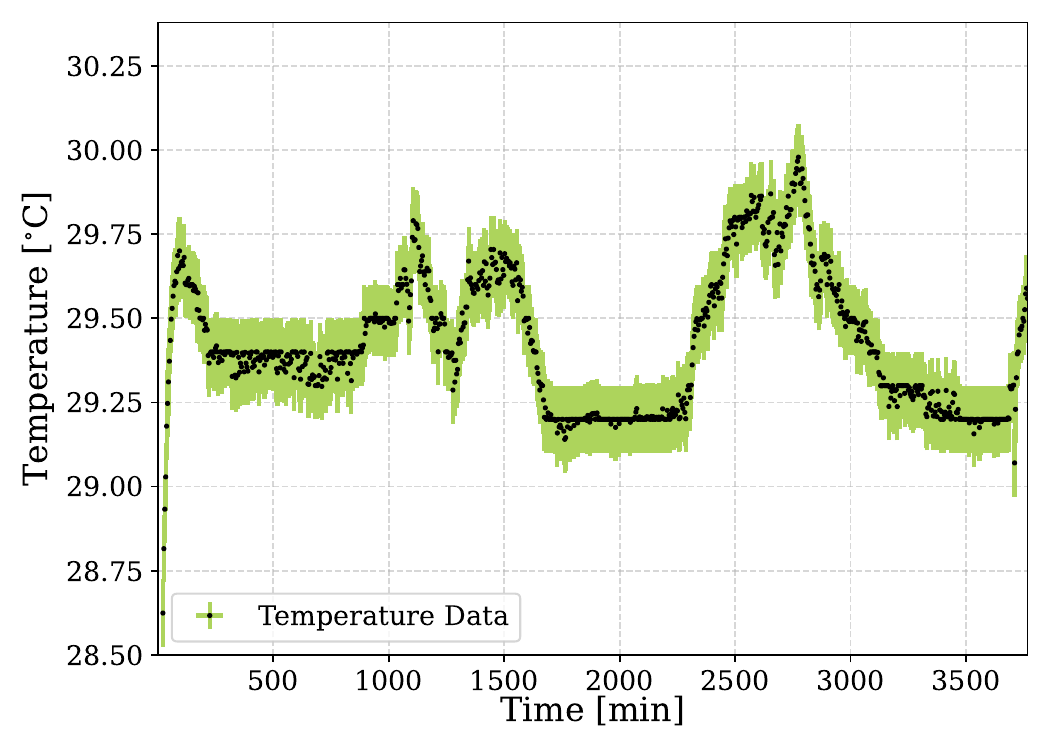}
        \label{fig:temperature}
    \end{subfigure}
    \caption{Left: Barometric pressure recorded by the onboard BMP280 sensor over time. Right: Temperature measured by the same sensor, used to monitor environmental conditions and correct for temperature dependent SiPM gain variations.}
    \label{fig:press}
\end{figure}

Figure~\ref{fig:mip_spectra} (right) displays the event rate versus time for the same dataset, comparing non‑coincident and coincident counts. Each point represents a 240\,s interval, with the rate calculated as the number of events divided by the effective live time (bin duration minus total dead time). These rate measurements enable searches for correlations with atmospheric variables; for e.g., barometric pressure changes (figure~\ref{fig:press} (left)) or transient phenomena such as solar flares. The detector also logs temperature (figure~\ref{fig:press} [right]), allowing corrections for temperature‑dependent gain shifts, although it currently isn't implemented.

\newpage
\subsection{Cosmic-ray muon angular dependence}\label{sec:angle}
\begin{figure}[b]
    \centering
    \begin{minipage}{.65\textwidth}
        \centering
        \includegraphics[width=\linewidth]{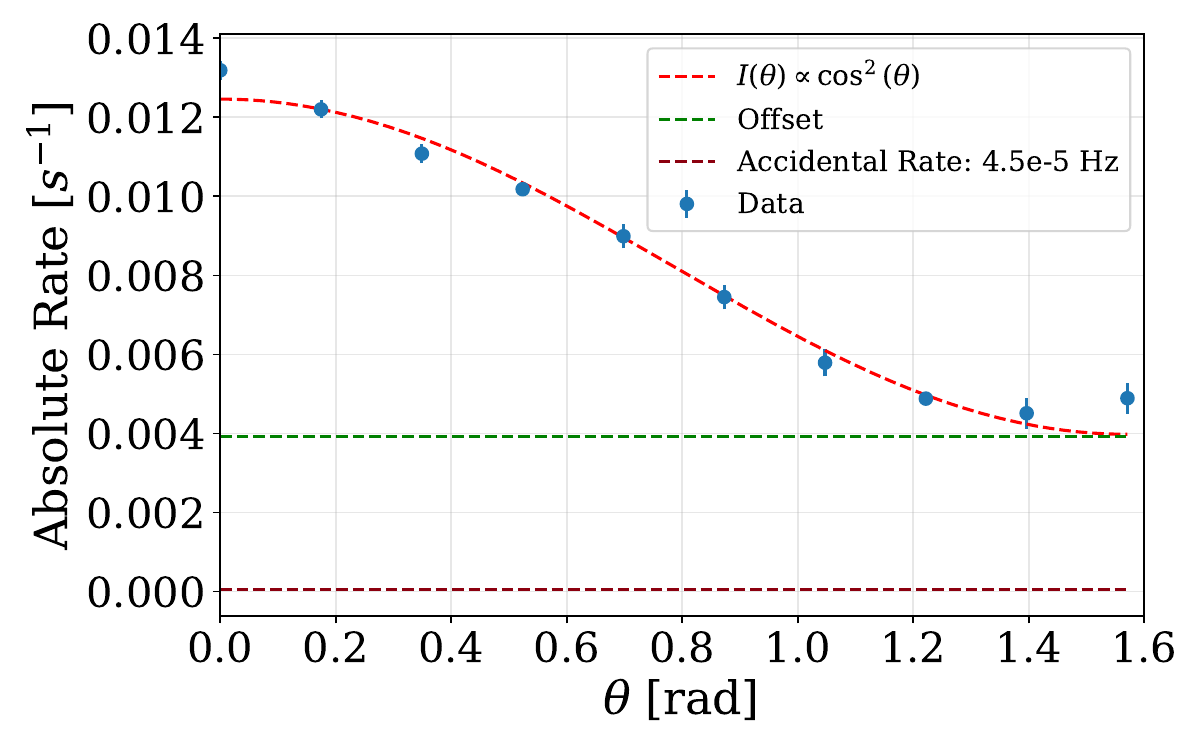}

    \end{minipage}%
    \begin{minipage}{0.35\textwidth}
        \centering
        \includegraphics[width=\linewidth]{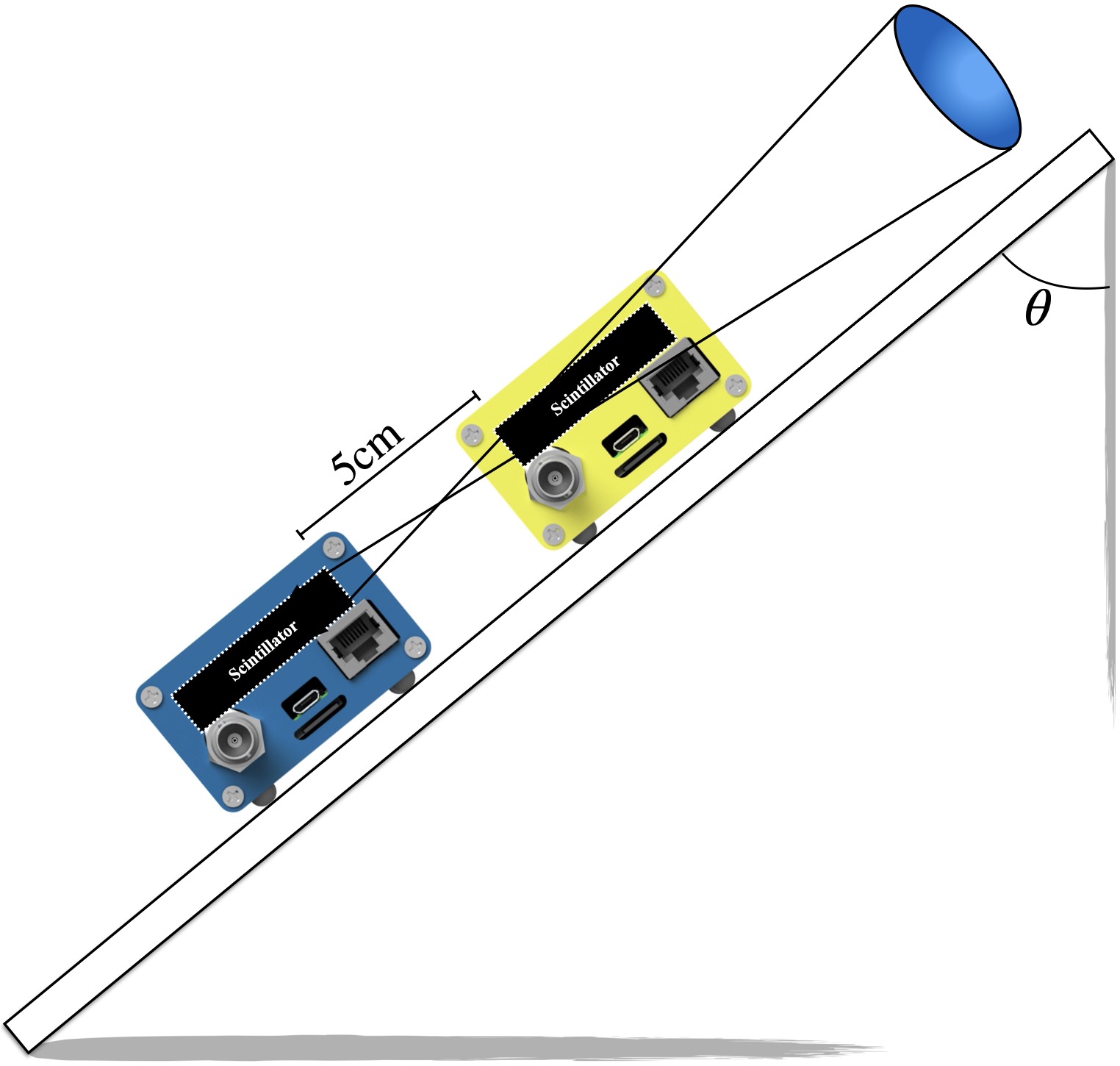}

    \end{minipage}
       \caption{Left: The measured cosmic-ray muon angular distribution along with the fitted components to the data. 
       Right: An illustration of the measurement configuration.}
       \label{fig::angle}
\end{figure}

This measurement illustrates the cosmic‐ray muon angular dependence. Two detectors were operated in coincidence mode and positioned side‐by‐side (configuration~(a) in figure~\ref{fig::opening_angle}), with their scintillators separated by 50\,mm inside identical aluminum enclosures. The detector angle relative to the zenith, \(\theta\), was set by mounting both units on a 40\,cm rectangular bar and leaning the bar against a vertical wall at a known height (see figure~\ref{fig::angle}, right). Figure~\ref{fig::angle} (left) presents the measured coincidence rate with statistical error bars, where each data point corresponds to roughly 24\,hours of data acquisition.

In this configuration, the zenith opening angle between the detectors was calculated to be approximately \(\pm8^\circ\). Several contributions to the measured rate are expected. First, the angular dependence of cosmic‑ray muons at sea level follows a cosine‑squared law, I$(\theta) \propto \cos^2 \theta$,
where \(I(\theta)\) is the differential muon intensity as a function of zenith angle \(\theta\) (with \(\theta=0^\circ\) corresponding to vertically downward‑going muons)~\cite{PDG_CosmicRays}. Second, there is an accidental coincidence rate, calculated as:
\begin{equation}\label{eq:accidental}
    R_{\rm acc} \;=\; 2\,N_1\,N_2\,\tau,
\end{equation}
where \(N_1 = 3.3\ \mathrm{Hz}\) and \(N_2 = 2.9\ \mathrm{Hz}\) are the individual detector trigger rates, and \(\tau = 2.3\ \mu\mathrm{s}\) is the coincidence window. Figure~\ref{fig::angle} (left) overlays the expected \(\cos^2\!\theta\) dependence and the flat accidental background rate. However, a residual offset remains in the data similar to that observed in Ref.~\cite{Axani_2018}. This excess is hypothesized to originate from components of shower development, perhaps due to knock-on electrons produced in the vicinity of the muon track.

The slight increase in rate at \(\theta = \pi/2\) is attributed to the detector configuration allowing acceptance of muons from both directions, unlike all other orientations, which record only downward-going muons. Consequently, the single-muon contribution at \(\theta = \pi/2\) should, in principle, be divided by two.

\newpage
\subsection{High-altitude balloon flight}\label{sec:hab}
Two detectors were flown in configuration (d) of figure~\ref{fig::opening_angle} on a high altitude balloon (HAB) to map the ionizing radiation flux as a function of altitude. The balloon ascended to 31\,km, deep into the stratosphere, where primary cosmic rays begin interacting with atmospheric nuclei to produce extensive particle showers. Above roughly 5\,km, the down-going ionizing flux of high energy ($\gtrsim$\,GeV) charged particles is dominated by protons~\cite{PDG_CosmicRays}. 

\begin{figure}[h]
\centering
\includegraphics[width=1\textwidth]{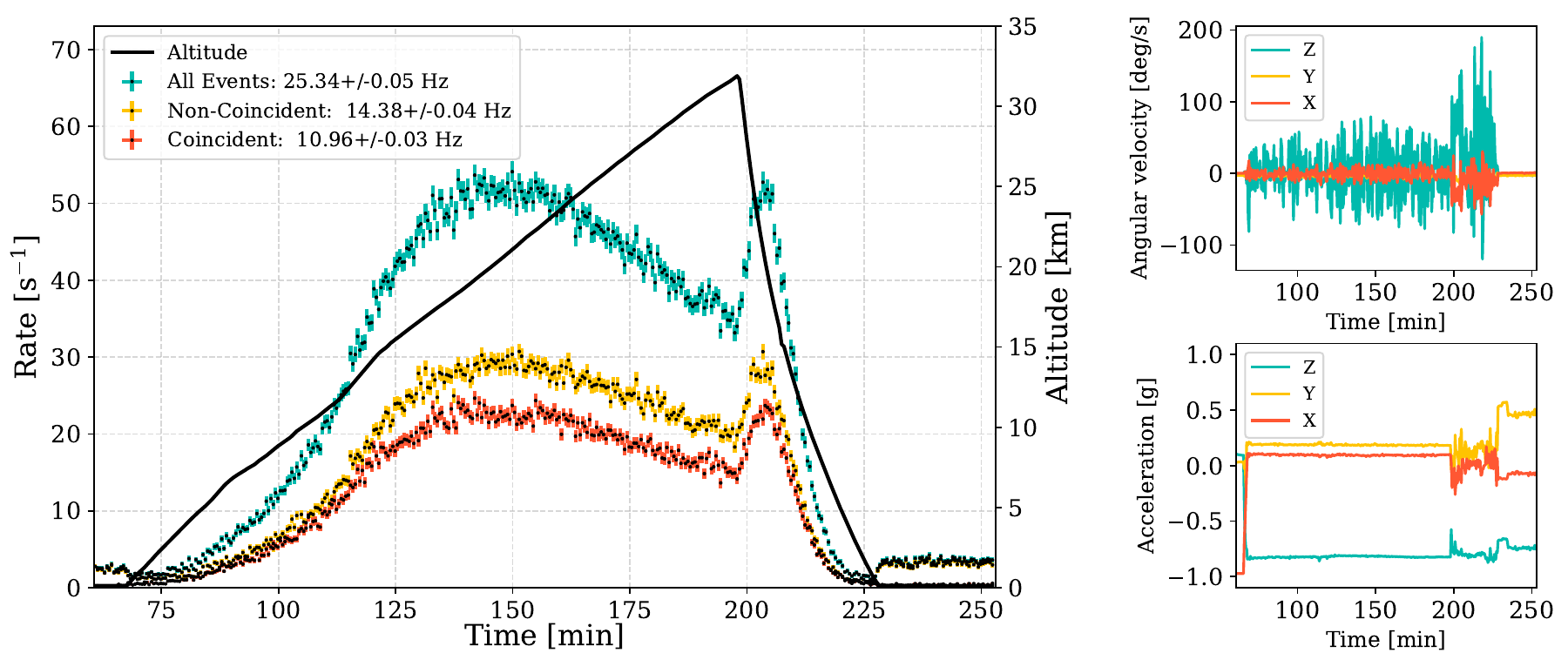}
\caption{Left: The measured detection rate as a function of time during a HAB flight. The right axis shows the altitude of the HAB, calculated from the onboard pressure sensor. Right: The measured angular velocity (top) and linear acceleration (bottom).}
\label{fig:hab_data}
\end{figure}

The detectors were powered by a single 5,000\,mAh battery pack, and data were recorded continuously to microSD cards. The onboard pressure measurements P was used to convert to geometric altitude h~\cite{united1976us}:
\begin{equation}
h \;=\; \frac{T_{0}}{L}
\left[
1 \;-\;
\Bigl(\tfrac{P}{P_{0}}\Bigr)^{\frac{R\,L}{g\,M}}
\right],
\end{equation}
where \(T_0 = 288.15\;\mathrm{K}\), \(L = 0.0065\;\mathrm{K\,m^{-1}}\), \(P_0 = 1013.25\;\mathrm{hPa}\), \(R = 8.31\;\mathrm{J\,mol^{-1}\,K^{-1}}\), \(g = 9.81\;\mathrm{m\,s^{-2}}\), and \(M = 0.029\;\mathrm{kg\,mol^{-1}}\). 
Figure~\ref{fig:hab_data} shows the event rate as a function of time (left axis), along with the calculated altitude (right axis). The data exhibit the characteristic altitude profile: count rate increases with altitude, reaching a broad maximum near 20\,km, the Regener–Pfotzer maximum~\cite{regener1935vertical}, where secondary cosmic-ray production and attenuation balance. Above this altitude, the flux decreases as diminishing atmospheric depth reduces secondary particle production. 

The built-in accelerometer indicated that the payload was initially oriented sideways during launch but realigned vertically shortly after liftoff (figure~\ref{fig:hab_data} lower right), and the tether remained stable throughout the flight. The gyroscope revealed significant rotation about the zenith throughout the flight (figure~\ref{fig:hab_data} upper right). The two detectors were mounted in a polystyrene foam enclosure, with a Mylar heat blanket. The temperature remained between 45$^{\circ}$C  (during lift-off and landing) and 23$^{\circ}$C (at apogee).

\subsection{Cosmic-ray muon velocity}\label{sec:velocity}

This measurement demonstrates the capability of BNC readout connected to the oscilloscope to perform time-of-flight (TOF) measurements to estimate the cosmic-ray muon velocities. Two detectors were vertically separated by 2.97\,m $\pm$\,0.01\,m, with their BNC outputs connected through equal length cables to a 50$\Omega$-terminated, 200\,MHz, 2.5\,GSa/s Tektronix MDO34 oscilloscope. The oscilloscope was used to save the SiPM pulse waveforms directly to a computer when both channels crossed a 5\,mV threshold. These waveforms were analyzed offline to extract precise pulse arrival times by fitting each waveform with a predefined pulse template. The time difference between pulses, \(\Delta t\), was then used to calculate the muon velocity using the relation v$ = d/ \Delta t$, where \(d\) is the detector separation.

\begin{figure}[ht]
    \centering
    \includegraphics[width=1\textwidth]{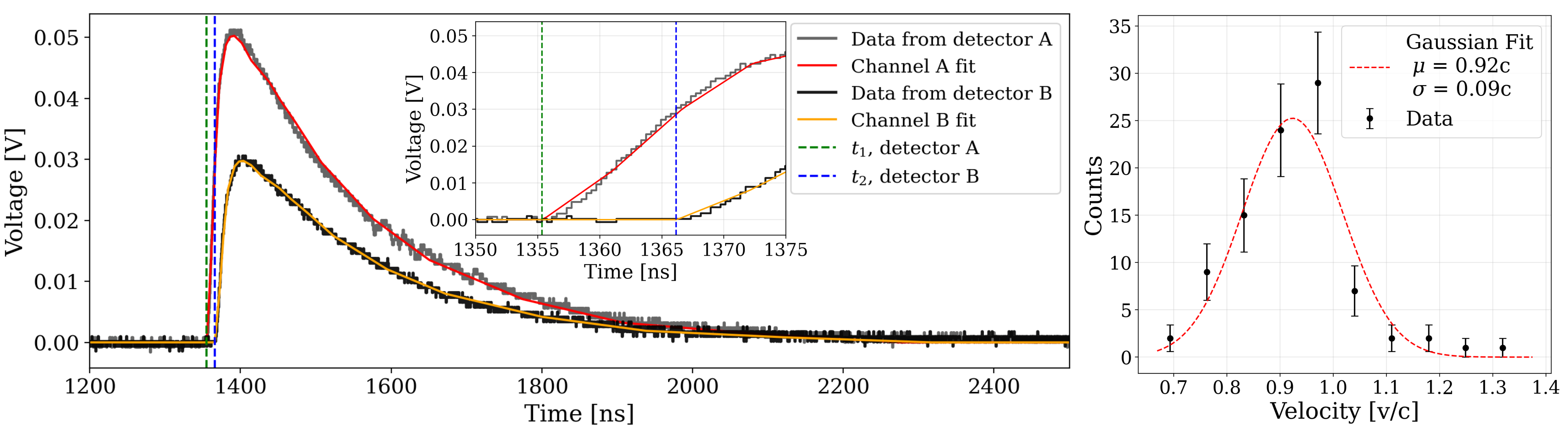}
    \caption{Left: An example waveforms and fitted pulses for top (red) and bottom (orange) detectors. Vertical dashed lines mark the muon arrival times determined using the pulse template. The inset image shows the rising edge of each pulse along with the fit. Right: The distribution of muon velocities, along with a fitted Gaussian in red.
    }
    \label{fig:muon_nu}
\end{figure}

The TOF measurement shown in figure~\ref{fig:muon_nu} recorded 92 events over the course of several weeks with a pulse separation of less than 10\,ns. A Gaussian fit to the resulting velocity distribution yielded a mean muon velocity of $0.92 \pm 0.09\,c$, where $\mu$ and $\sigma$ correspond to the mean and standard deviation, respectively. The expected number of accidental coincidence events, calculated using equation~\ref{eq:accidental}, was found to be less than one over the duration of the measurement.

\section{Conclusion}

The CosmicWatch Desktop Muon Detector v3X represents a significant advancement in compact, low-cost particle detection. With redesigned analog electronics, a dual-core MCU architecture, and enhanced digital features, the v3X achieves precise, high-rate event acquisition while remaining accessible to students and educators. Key upgrades including improved timing resolution, lower noise, sub-millisecond dead time per event, and native support for coincidence triggering; enable new measurements such as muon time-of-flight, energy deposition studies, and environmental correlations.

Despite these enhancements, the detector retains its core strengths: affordability, ease of assembly, and flexibility. It can be operated as a standalone unit for classroom experiments or integrated into large-scale educational and citizen-science networks. Demonstrated applications range from gamma-ray spectroscopy to high-altitude radiation profiling and cosmic-ray muon tracking.

Through this work, we aim to lower the barrier to experimental physics by providing a versatile, modern detector platform that can serve as both a research instrument and an educational tool.


\section{Acknowledgments}

The authors gratefully acknowledge Amanda Meng, Liam Roth, Maeve Owens, Musarate Shams, Michael Manfre, Christina Love, Rafael Hurtado Carrillo, and Andres Felipe Vargas for their assistance with the measurements presented in this work. We also thank our colleagues at MIT, WIPAC, and Harvard for their support and for helping to integrate this effort into a high‑school and undergraduate research initiative.


\appendix
\section*{Appendix}
\addcontentsline{toc}{section}{Appendix} 

\section{Electronics Schematic}

\begin{figure*}[ht!]
\centering
\includegraphics[width=1.40\textwidth, angle=90]{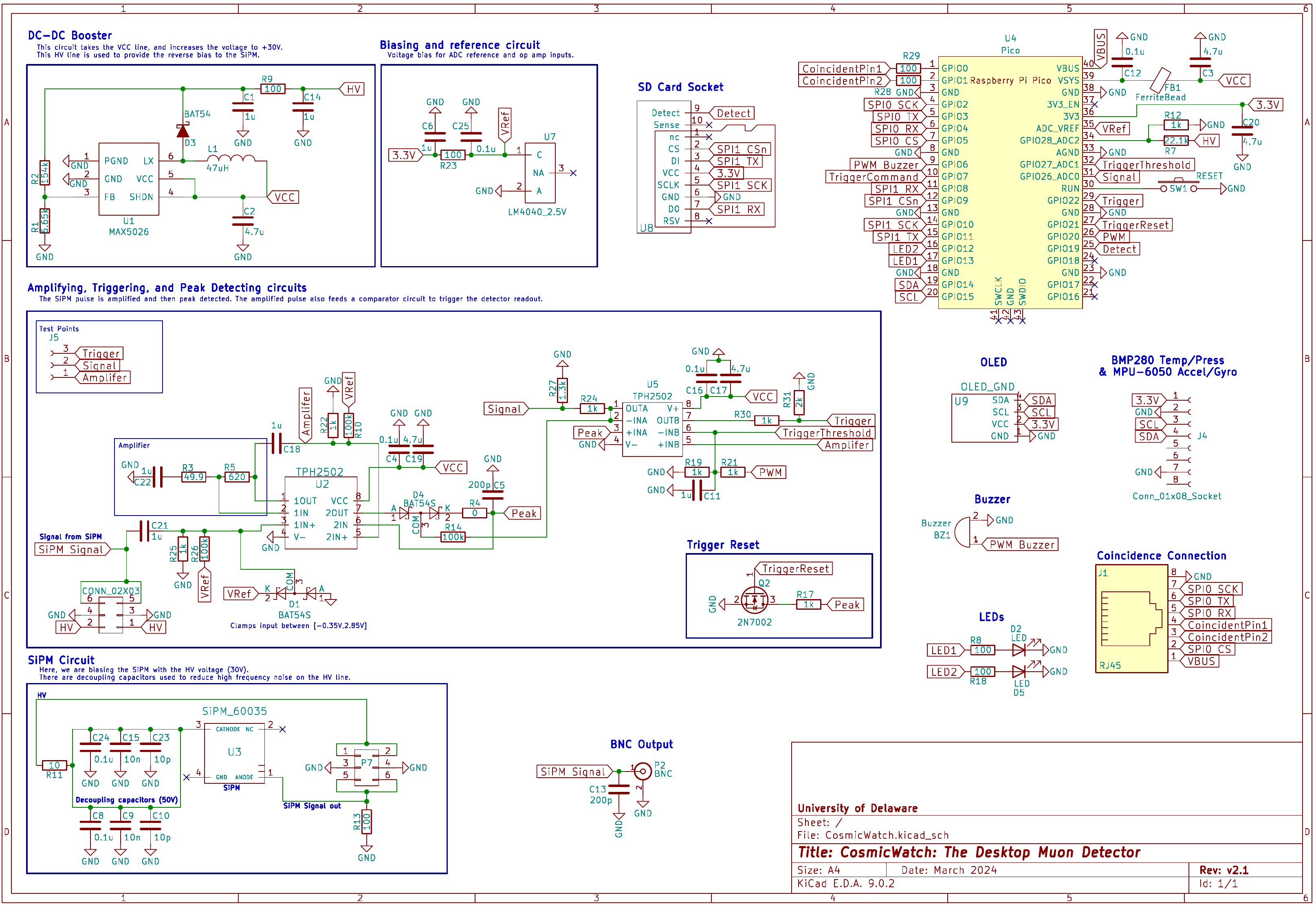}
\caption{Complete detector electronic schematic. }

\label{fig:schematic}
\end{figure*}

Figure~\ref{fig:schematic} shows the full electrical schematic of the detector system. The design integrates the SiPM, power regulation, analog signal processing, triggering logic, MCU, and microSD storage onto a compact, 2-layered PCB layout. The SiPM is biased via a precision low-noise DC to DC booster, and its output is fed into a high-speed rail-to-rail operational amplifiers followed by a comparator that defines the trigger condition. The trigger logic initiates data acquisition, which are managed by the RPi running custom firmware. Digital data are stored to a microSD card via SPI communication. The design also incorporates decoupling capacitors, protective elements to ensure robust operation in noisy environments. This compact integration allows for a self-contained, low-power, and portable particle detector system suitable for field use, classroom environments, or as part of a larger detector array.

\newpage

\bibliographystyle{JHEP}
\bibliography{biblio} 
\end{document}